\documentclass[prl,aps,twocolumn,showpacs]{revtex4}
\usepackage[fleqn]{amsmath}
\usepackage{amsfonts}
\usepackage{psfrag}
\usepackage{graphicx}
\usepackage{graphics}
\usepackage{bm}
\usepackage{color}
\usepackage[normalem]{ulem}

\newcommand{\be}{\begin{equation}}
\newcommand{\ee}{\end{equation}}
\newcommand{\ba}{\begin{eqnarray}}
\newcommand{\ea}{\end{eqnarray}}

\newcommand{\dcom}[1]{}
\newcommand{\dnote}[1]{}

\newcommand{\gsim}{\raise.3ex\hbox{$>$\kern-.75em\lower1ex\hbox{$\sim$}}}
\newcommand{\lsim}{\raise.3ex\hbox{$<$\kern-.75em\lower1ex\hbox{$\sim$}}}

\begin{document}

\title {Large phonon time-of-flight fluctuations in expanding flat condensates of cold fermi gases}

\author{Da-Shin\ Lee  and Chi-Yong\ Lin}\affiliation{
Department of Physics, National Dong Hwa University, Hua-Lien,
Taiwan 974, R.O.C.  } \author{Ray\ J.\ Rivers}
\affiliation{ Blackett Laboratory, Imperial College
London, SW7 2BZ, U.K.}

\date{\today}
\begin{abstract}
We reexamine how quantum density fluctuations in condensates of ultra-cold fermi gases lead to fluctuations in phonon times-of-flight, an effect that increases as density is reduced. We suggest that these effects should be measurable in pancake-like (two-dimensional) condensates on their release from their confining optical traps,  providing their initial (width/thickness) aspect ratio is suitably large.
\end{abstract}


\maketitle
\section{Introduction}
We are familiar with the idea that the quantum nature of the gravitational field induces fluctuations in the metric of space-time (the light-cone metric), thereby inducing fluctuations in the propagation of light \cite{ford}.
What is less familiar is that a similar situation applies to bosonic condensates of ultra-cold fermi gases whose quantum fluctuations disturb the sound-cone metric, leading to fluctuations in the time-of-flight (TOF) of phonons and sound-waves.

The reason is straightforward. Whereas, for a condensate of elementary bosons, there is only the gapless phonon mode, for condensates of fermi gases di-fermion density fluctuations act as an additional effective gapped mode \cite{Lee6}. This decoheres the phonon, giving the sound-cone metric a stochastic component \cite{Lee3}. { There are parallels with the effect of dark energy on gravitational waves \cite{de_rham}} as well as the simpler case of photons \cite{hu,ford} and phonons \cite{gurarie,krein,gaul} in phenomenological random media except that fluctuations in fermi-gas sound-propagation (like quantum fluctuations in the propagation of light) are determined by the same interactions that create the phonons themselves.

Quantitatively there are huge differences between photons and phonons, with light-cone fluctuations being immeasurably small by many orders of magnitude \cite{ford,de_rham}, whereas sound-cone fluctuations are relatively large. However, in a previous paper \cite{Lee3} we estimated the stochastic effects on phonon TOF for a typical three-dimensional static homogeneous condensate and concluded that they would still be experimentally unobservable by something between one to two orders of magnitude. In particular, TOF fluctuations decrease with increasing condensate density and the density of typical condensates is too large for measurable effects.

In this paper we exploit this density effect by calculating TOF fluctuations as we release a condensate at $T\simeq 0$ from its  {\it optical} trap, with a rapid decrease in its density.  In addition, we differ from our previous analysis in taking our condensate  to be in  a
highly anisotropic pancake-shaped potential initially~\cite{exp1,exp2,exp3,exp4,ries,supp}. Not only does this permit the density profile of the effectively two-dimensional cloud  to be measured {\it in situ} but lower dimensionality enhances fluctuations further. We argue that the resulting TOF fluctuations may now be large enough to be measurable with current techniques.

In
recent years several papers have been devoted
to the study of quantum/thermal fluctuations in such idealised 2D gases ~\cite{theory1,theory2,theory3}. See~\cite{review} for a more general study.
In particular, the BCS-BEC crossover of two-dimensional Fermi gases
was studied in~\cite{hu1}
where the inclusion of  quantum fluctuations about the mean field theory found good agreement with the quantum Monte Carlo simulations and the experimental measurements. In this work, we will adopt the same level of approximation. The fluctuations we consider here are purely quantum, for an idealised gas at temperature $T=0$. The analysis requires some care since, on releasing the trap, the initially flat condensate will expand (approximately) from a disc to a cylinder, rapidly ceasing to be two-dimensional. It is possible to provide some interpolation between D=2 and D=3 dimensions (e.g. see \cite{theory3,valiente,olaussen}). We are less ambitious in that we shall restrict ourselves to calculating the metric fluctuations only from the period of expansion when the condensate can be considered two-dimensional when the effect of quantum fluctuations is larger. Although this is an under-estimate the metric fluctuations from the three-dimensional regime are generically smaller and the effect of ignoring them is not large.

There is a potential problem in that, for two-dimensional quantum systems finite temperature effects will destroy the true long-range coherence due to large long-wavelength phase fluctuations even in the superfluid states. However, a transition to a superfluid state with quasi-long-range-order (the Berezinskii-Kosterlitz-Thouless (BKT) transition) will occur when the temperature is below some critical temperature $T_{\rm BKT}$. This superfluid state  has been experimentally observed in  ultracold fermi gases across BEC-BCS crossover~\cite{mur}.
The experiments that we have in mind  for successfully observing pair condensates in this quasi-2D system are those in~\cite{ries,mur,strecker} at temperatures lower than  $T_{\rm BKT}$ for which, at low enough temperature, sizable condensates of the order $100 \mu m$ can exist. {The expansion of the condensate lowers the temperature further.

To describe our quasi two-dimensional state we dimensionally reduce the spatially 3D action that has been the basis of our earlier work \cite{Lee3,Lee1,Lee2,Lee_3,Lee4} to the 2D action. This action, due to Gurarie~\cite{gurarie2}, describes a cold
 Fermi gas, tunable through a {\it narrow} Feshbach resonance, taking the form (where suffices denote spatial dimension):
\begin{eqnarray}
S_{(3)} &=& \int dt\,d^3x\bigg\{\sum_{\uparrow , \downarrow}
 \psi^*_{(3) \sigma} (x) \left[ i \
\partial_t + \frac{\nabla^2}{2m} + \mu_{(3)} \right]  \psi_{(3) \sigma} (x)
\nonumber
\\
   &+& \varphi^{*}_{(3)}(x)  \left[ i  \ \partial_t + \frac{\nabla^2}{2M} + 2 \mu_{(3)} -
\nu \right]  \varphi_{(3)} (x)
\nonumber
\\
&{+}& g_{(3)} \left[ \varphi^{*}_{(3)} (x) \ \psi_{(3) \, \downarrow} (x) \ \psi_{(3) \,\uparrow}
(x) + \rm{h.c.}
\right]\bigg\}
\label{Lin}
\end{eqnarray}
%\vskip -0.3cm
%
\noindent
for fermion fields $\psi_{(3) \,\sigma}$
 with spin label $\sigma = (\uparrow, \downarrow)$. The diatomic field $\varphi_{(3)}$ describes the bound-state (Feshbach) resonance with
 tunable binding energy $\nu$ and mass $M =2m$.
 %To keep the algebra as simple as possible we ignore direct four-fermi interactions and {\it broad} Feshbach resonances~\cite{%path_integral1,
 %path_integral2,path_integral3,path_integral4,path_integral5}. We shall introduce them later.

The condensate can be reduced to D=2 dimensions if
its thickness, say along the $z$ axis, is tightly confined
by an external {optical} trapping  potential to be no thicker than
the correlation length of the system $L_z\sim l = 1/mc$, where $c$ is the speed of sound. As a
consequence,  the fields of the system are approximately constant along the
$z$ direction and on dimensional grounds they can be simplified
as
$
\varphi_{(3)}({\bf x}_\bot,z,t)= {\varphi}_{(2)}( {\bf x}_\bot,t)/L_z^{1/2} \, ,
\psi_{(3)}({\bf x}_\bot,z,t)= {\psi}_{(2)}({\bf x}_\bot,t)/L_z^{1/2}$,
where ${\bf x}_\bot=(x,y)$.
Consequently, the effective coupling ${g}_{(2)}$ in two dimensions
is related to that of three dimensions by
$ g_{(2)}=g_{(3)}/L_z^{1/2}$.   As such, the spatial integral over $z$ can be carried out,
and the resulting action $S_2$ of (\ref{Lin}) for D=2 has no explicit $L_z$ dependence.

%We shall need both $S_{(2)}$ and $S_{(3)}$ since,
On releasing the trap the condensate extends in the direction perpendicular to the pancake, ultimately losing its two-dimensional form. Quantitatively we shall show that the dominant contribution to the TOF fluctuations occurs when the condensate is relatively flat, diminishing as it becomes fully 3-dimensional. It is sufficient to evaluate the fluctuations in the 2-dimensional regime when the dimensional reduction is appropriate.

{It need hardly be said that, although we were motivated by analogy with quantum fluctuations in the gravitational and light-cone metrics, the subsequent analysis of fluctuations in the sound-cone metric of ultracold gases stands in its own right.}

\section{The effective theory}

All that follows is for the 2-dimensional theory. Henceforth we shall drop the suffixes denoting 2-dimensionality unless otherwise specified, assuming that we are below the BKT transition where such low temperature allows to ignore thermal fluctuations. Since the 2D reduced action (\ref{Lin}) is quadratic in the Fermi fields, they can be integrated out to give an exact {\it non-local} (one fermi-loop) bosonic action $S_{NL}[\varphi, \varphi^*]$ \cite{Lee3}.
Initially we shall take the condensate to be homogeneous and in equilibrium. The action $S_{NL}$ is invariant under global $U(1)$ transformations of $\varphi$ but the variational equation $\delta S_{NL} = 0$ permits constant condensate solutions $\varphi (x) = \varphi_{0}$, which break this symmetry spontaneously, determined by the gap equations (see Appendix).

To relate the condensate parameters in D=2 and D=3 dimensions we use the fermi momentum $k_F$ as the interpolating dimensionally-dependent parameter~\cite{hu2,hu3}, introduced by defining the effective column density
$\rho_{(2)}$ in effective 2D in terms of the density in 3D $\rho_{(3)}$ by $\rho_{0 (2)}=\rho_{0(3)} L_z$. Both the {\it constant} 2D fermion number density  $\rho_{0(2)} =k_{F(2)}^2/2 \pi$ and the 3D fermion number density $\rho_{0(3)} =k_{F(3)}^3/3\pi^2$ are decomposable as $\rho_{0} = \rho^F_{0} +\rho^B_{0} $, where $\rho^F_{(0)}$ is the explicit fermion density and $\rho^B_{(0)}$
is the molecule density (two fermions per
molecule).

The interpolating scattering parameter for BCS-BEC crossover is $u =1/{a_{s} k_{F}}$ where $a_s$ is the s-wave scattering length. D=3 and D=2 dimensions show very different general behaviour. For D=3 there is a unitary point $u = 0$ which separates the BCS regime ($u < 0$) from the BEC regime ($u > 0$). The deep BCS regime and deep BEC regimes are $ -u\gg 1$ and  $ u\gg 1$ respectively. In D=2 dimensions $u > 0$ always and the deep BCS regime is $u\ll 1$ whereas the deep BEC regime is $u \gg 1$ as before. We are primarily interested in the deep BEC regime where there is no ambiguity.

As expected, in the deep BEC regime $\vert \varphi_{0} \vert^2 \rightarrow \rho_{0}/2$
with $\rho^B_{0} \gg \rho^F_{0}$.
The condensate of the theory  is
$\varphi (x) = |\varphi(x)| \ e^{i
\theta (x)}$
with the gapless mode encoded in the phase $\theta$.
 We expand  in the derivatives
of $\theta$ and the {\it small} fluctuations in the condensate
 $\delta\varphi = (|\varphi|  -  |\varphi_{0}|) \propto \delta\rho^B$, the molecular density fluctuation, explicitly preserving Galilean invariance.
 The {\it local} effective
action for the long-wavelength, low-frequency condensate has  the
same generic form in terms of $\theta$ and $\delta
|\varphi|$ in two as in three dimensions \cite{Lee3}, but with coefficients appropriate to the dimension
\cite{Lee1,Lee2}:
\begin{eqnarray}
 S_{\rm eff}[\theta, \epsilon] &=& \int dt~d^2x\bigg[ \frac{N_0}{4}\ G^2(\theta, {\epsilon}) -\frac{1}{2}{\rho}_0
   G(\theta, {\epsilon})
 \nonumber
 \\
 &-&{\alpha}{\epsilon}G(\theta, {\epsilon})
 +\frac{\eta}{4}{}D_t^2({\epsilon},\theta)
  -\frac{1}{4}{\bar M}^2{\epsilon}^2\bigg],
 \label{LeffU0}
 \end{eqnarray}
\noindent
 where $\epsilon = \kappa^{-1}\delta|\phi|$
is a {\it dimensionless} rescaled condensate fluctuation.  The scale factor
is chosen so that the Galilean scalar $G(\theta, \epsilon ) = \dot{\theta} + (\nabla
\theta )^2/4m + (\nabla \epsilon )^2/4m$ is such that $\epsilon$ has the
 same coefficients as $\theta$ in its spatial derivatives \cite{aitchison}.  $D_t$ is the comoving
time derivative in the condensate with fluid velocity $\nabla
\theta/2m$. For ease of notation we have not labelled the fields and their coefficients by their dimension explicitly. We shall  omit suffices when it leads to no confusion.

Sufficient to say that,  {in both D=2 and D=3 dimensions}, $0\leq
\alpha/\rho_0$ increases monotonically from zero to one as we tune the gas from the deep BCS
 to the deep
BEC regime ($1/k_F a_S \gg 1$). Similarly $\eta$ and $N_{0}$  go from finite values to zero
as we go from the BCS to BEC regimes, with  $\eta\propto
N_{0}$ in the deep BEC regime (as they both vanish). Further, ${\bar M}^2$ falls to zero in the deep BEC regime
 in a way that ${\bar M}^2/\eta$ remains finite. In D=2 dimensions the coefficients $\alpha, \eta$, etc. are known elementary functions of the scattering length \cite{Lee1,Lee2}.  See the Appendix  below.

We only know how to model the condensate expansion within the hydrodynamic approximation, where the spatial and temporal
variation  of $\epsilon$ are ignored in comparison to $\epsilon$
whence $\epsilon \approx -2 \alpha G(\theta)/{\bar M}^2$. We know from elsewhere \cite{Lee3} that the approximation breaks down badly outside the BEC regime  and we show that we can stay within it.  The
  Euler-Lagrange equation for $\theta$
 is then the continuity equation of a {\it
single} fluid \cite{Lee1,Lee2}
\be
\frac{\partial}{\partial t} \rho+\nabla \cdot (\rho {\bf v})=0\, \label{cont}
\ee
\noindent
with $\rho=\rho_0-(N_0 + 4 \alpha^2/{\bar M}^2) G(\theta)$ and ${\bf v}=\nabla\theta/2m$.
For long wavelengths it leads to the wave
equation:
\be
 \ddot \theta (x) -c^2  \nabla^2\theta (x)=0 \,
 \ee
 i.e. $\omega^2 = c^2k^2$, with speed of sound
\ba
c^2&=&\frac{\rho_0/2m}{N_0+ (4\alpha^2/\bar{M}^2)} \,
\label{c}
\ea
a result unchanged in the Bogoliubov approximation \cite{Lee6}. In the deep BCS regime $u\approx 0$, $c^2 \rightarrow v_{F(2)}^2/2$, as expected, but a particular feature of this simple model is that $c \rightarrow 0$  in the deep BEC regime. Specifically, if $\bar g^2\ll 1$ is the dimensionless coupling strength (defined later) then, in the BEC regime ($u^2 \gg {\bar g}^2)$
 we expand $c^2$ of (\ref{c2})  in powers of ${\bar g}^2$ (see Appendix). This gives, to leading order,
\be
{c =\frac{v_F}{\sqrt{2}}\frac{{\bar g}^2}{u^2}.}
\label{c2approx}
\ee
[The speed of sound also vanishes in the deep BEC regime for D=3 dimensions in the same approximation.]

It is not difficult to change this to give a non-zero limit by taking the direct dimer-dimer interaction into account~\cite{Lee6},  but, since we never achieve the really deep BEC regime, we take it as it stands.
More generally,  there is a quantum 'rainbow' of sound speeds
$c_k$ \cite{Lee6,jain2,jain3}, according to the wavenumber $k$ of the phonon, of the form \cite{Lee3}
$ c_k^2 \approx  c^2[ 1 + k^2/K^2 + ...]$
where
 $ K^{-2} = 4\alpha^2 c^2\left[ 1- c^2 \eta/(\rho_0/2m) \right]/{\bar M}^4 $. {[We note that the 'gravitational rainbows' due to dark energy fields take the same functional form \cite{de_rham}.]}
 In the large momentum limit in the BEC regime we recover \cite{Lee2} the free particle limit for diatoms/molecules
$\omega = k^2/4m = k^2/2M$. Provided that the phonons comprise a wavepacket propagating
 on the plane with  central momentum $k_0$ and width $\Delta k_0$, with $k_0 + \Delta k_0 < K$,
they experience approximately the same sound speed $c$ and have
common fluctuations in times of the flight, and this we assume.

To complete the fluid picture we observe that the above definition of $\rho$ is no more than the Bernoulli equation
\be
m \dot{\bf v}+\nabla \big[ \delta h + (1/2)m v^2 \big]=0 \, , \label{BLeq}
\ee
\noindent
where the enthalpy $\delta h =m c^2 \delta \rho/\rho$. The resulting equation of motion is $ d p/d \rho= m c^2$ across the whole regime for which the hydrodynamic model is appropriate.
Since $c^2 \propto \rho$, there is simple allometric behavior $\delta h \propto \rho^\gamma$ with $\gamma=1$~\cite{Lee1,Lee2}, from which the hydrodynamic behavior of the fluid can be determined.

\section{The stochastic metric}

The semiclassical results above do not take the dynamical
nature of the density fluctuations into account. Because Galilean invariance induces {\it multiplicative} noise~\cite{Lee3}, the coarse-graining
of the phonon field induced by integrating out $\epsilon$ will introduce stochasticity in the
acoustic metric of the $\theta$ field via its Langevin equation rather than dissipation. Specifically, in the long-wavelength approximation the stochastic equation for phonon propagation
now becomes  \cite{Lee3}
 \be
 {\ddot\theta}(x) - c^2(1 - 2\alpha\xi/\rho_0)\nabla^2\theta\approx  - 4m(\alpha/\rho_0) c^2 {D_t \xi}(x)\,
 \label{langac2}
 \ee
\noindent in terms of the speed of sound $c$ of (\ref{c}) and noise $\xi$. As a result we can
interpret $c_{\xi}$, where $c_{\xi}^2 = c^2 (1 - 2\alpha\xi/\rho_0)$,
 as a stochastic speed of sound in the long wavelength regime.

 As we have shown in \cite{Lee3} for D=3, but which generalises to D=2 dimensions, the noise correlation function is the Hadamard function for the density fluctuation field $\epsilon$,
\ba
\langle\xi(x)\xi(x')\rangle &=& \frac{1}{2} \langle \{ \epsilon(x), \epsilon(x') \}
 \rangle
 \nonumber
 \\
  &=&
 \int \frac{d^2{\bf k}}{( 2\pi)^2}
 \frac{\cos[{\omega}_k (t-t')]}{{\omega}_k {\eta}} \, e^{-i {\bf k} \cdot ({ \bf x}-{\bf x'})}.
 \,
 \label{D_H}
\ea
with $\omega_k^2 = (\rho_0 k^2/2m +\bar M^2)/\eta$.

 This effect can be tested by the
variation in the TOF of the phonons or sound waves traveling in a fixed distance, by repeated measurements of the sound speed of the identical wavepackets.
The experiment that we have in mind is along the lines of~\cite{joseph}, which measures   the speed of sound  in a fermi gas  throughout the crossover region, however in the case of broad resonances. Such an experiment with Feshbach resonances would be technically more challenging~\cite{strecker} even though narrow resonances lead to relatively stable and long-lived molecules.

For a spatially homogeneous static condensate the sound wave propagates along the sound
cone determined by its operator-valued null-geodesic:
\be
c^2 d t^2 = d {\bf{x}}^2 +h_{ij} dx^i dx^j \, ,
\ee
where $h_{ij} = (2\alpha/\rho_0)\xi\delta_{ij}$.
 For waves travelling a distance $r$ in time $T$, the variation of the travel time in two spatial dimensions is
\be
(\Delta T)^2 = \!\! \frac{\alpha^2}{\rho_{0(2)}^2}\!\! \int_0^{ T
} d t_1 \int_0^{ T } d t_2 \langle \xi (r_1,t_1) \,
\xi(r_2,t_2) \rangle  \, .
\label{Dt}
\ee
and we have restored the suffix to $\rho_0$.
\noindent
In (\ref{Dt})
$r(t)=|{\bf x}|= c t$ and the local velocity $c$ is evaluated on the unperturbed path of
the waves $r(t)$.

Straightforward substitution of
(\ref{D_H}) gives
\ba
\! \left(\Delta T \right)^2 \!
= \! \frac{\alpha^2}{\rho_{0(2)}^2} \!\int_0^T\! \!
dt_1\!\int_0^T \!\!  dt_2 \!\!
\int  \! \frac{d^2 k}{(2\pi)^2}~
 \frac{\cos({\omega}_{k} \tau)}{{\omega}_k \eta} \nonumber\\
 \times J_0 (c k \vert \tau \vert ) \nonumber\\
\label{deltat/t_exact}
\ea
\noindent
where $\tau = t_1 - t_2$ and $ J_0 $ is the Bessel function of the first kind. The initial growth
of $\left(\Delta T \right)^2$ from zero at time $T=0$ is a consequence of the accumulation of the zero temperature quantum fluctuations of the gapped modes. Inspection shows that when $T \sim T_s = 1/\omega_{(k=0)}
= 1/\sqrt{\bar{M}^2/\eta}$ (i.e. inversely proportional to the gap energy of the gapped (Higgs) modes),
the growth halts and $\left(\Delta T \right)^2$  saturates to its late time value $\left(\overline{\Delta T }\right)^2$.

In more detail assume that $\omega_k\gg c k$ throughout the range of the BEC and BCS regimes and take
$J_0 [ck \vert t_1-t_2 ] \approx 1$ (as compared with the relatively fast varying time-dependent cosine function). As a result,
\ba
 \left(\Delta T \right)^2
&\approx&   \frac{\alpha^2}{\rho_{0(2)}^2} \, \frac{2}{\eta}
\int  \frac{dk~ k}{(2\pi)}
 \frac{1-\cos[{\omega}_{k} T]}{{\omega}_{k}^3}\nonumber
 \label{deltat/t_approx}
 \ea

For $T\gg T_s$ the oscillatory term can be ignored and the fluctuations saturate to the value \be
 \left(\overline{\Delta T} \right)^2 \approx \frac{\alpha^2}{\rho_{0(2) }^2} \, \frac{2 m}{\pi \rho_{0(2) }} \frac{1}{ ({\bar{M}}^2/\eta)^{1/2}} \, . \label{deltat2_s}
 \ee
 We note that $ \left(\overline{\Delta T} \right)^2$ is independent of T and hence the distance over which the phonons are propagated. This variation of the arrival time can be experimentally tested by repeatedly measuring the arrival time $T$, traveled by the sound wave for a fixed distance with identical wavepacket forms.

 If $L_{\bot}$ is the effective breadth of the condensate the time for a phonon to traverse it is ${\bar T}\approx L_{\bot} /c$. To identify fluctuations cleanly we require
 \be
 {{\bar T}^2 \gg  \left(\overline{\Delta T} \right)^2 \gg T_s^2,}
 \label{ggg}
 \ee
{with $\overline{\Delta T}$ large enough to be measurable. This is not straightforward to achieve.}

\section{$\overline{\Delta T}$ for static pancake condensates}

From our earlier comments we treat the condensate initially as 2D.
Substituting the 2D parameters (of the Appendix) into~(\ref{deltat2_s}) we find that $\left(\overline{\Delta T} \right)^2$ increases as $1/a_S k_F$ increases monotonically, to
 reach its asymptotic maximum value (in dimensionless parameters, restoring dimensional subscripts where necessary) in the BEC regime
 \be
 \left( \epsilon_{F(2)}\overline{\Delta T_{(2)}} \right)^2 \approx (1/ \sqrt{12}) {\bar g}_{(2)}.
 \label{deltat2max}
 \ee
  where ${\bar g}_{(2)}$ is the dimensionless coupling constant defined as $g^2_{(2)} = (8 \pi \epsilon_{F (2)}/m){\bar g}^2_{(2)}  =16 \pi \bar g^2_{(2)} (\epsilon_{F(2)}^2/k_{F(2)}^2)$.

 To see what this means for a typical static pancake condensate, consider a cold $^6Li$ condensate of $ 3\times 10^5
$ atoms tuned by the narrow resonance at $H_0 = 543.25 G$ \cite{strecker}.
The narrowness of the resonance width is best determined
by the dimensionless width
 $\gamma_0\approx\sqrt{\Gamma_0/\epsilon_{F (3)}}$, where the resonance width
$\Gamma_0$   \cite{gurarie2} is mainly given by $H_{\omega}$,
the so-called "resonance width"  of the central field $H_0$, which in turn determines the effective-range length of the system, and is required
to achieve infinite scattering length (the unitary limit).
We take the number density $\rho_{0 (3)}
\approx 3 \times 10^{12} cm^{-3}$, and $\gamma_0\approx 0.6$. In terms of the
dimensionless coupling $\bar g_{(3)}$, where $g_{(3)}^2 = (64\epsilon^2_{F (3)}/3
k_{F (3)}^3){\bar g_{(3)}}^2$ \cite{gurarie2}, $^6Li$ at the density above
corresponds to ${\bar g_{(3)}}^2 $ of order $0.8$.

The pancake-shaped trap potential of \cite{strecker}, which we adopt here,  has frequencies $\omega_x = \omega_y = \omega_{\bot}= 2\pi \times 0.05 Hz$ in the plane with characteristic length $L_{\bot 0} =480 \mu m$, and $\omega_z= 2\pi \times 800 Hz$ in the vertical plane for a thickness $L_{z 0}=1.4 \mu m$.  The trap anisotropy ratio is $\omega_z/\omega_{\bot} \approx 10^4$, allowing us to treat the system as quasi-2D dimensional for any speed of sound that is a fraction of $v_F$, since $1/mv_F\approx 1 \mu m$.
As a first approximation we ignore edge effects, relying on the relatively homogeneous central part of the condensate.
The density of this quasi-2D system can then be estimated from the 3D density by  $\rho_{0 (2)}= \rho_{0 (3)} \times L_z= 10^8 cm^{-2}$
for
which $\epsilon_{F(2)} \approx 1 \times 10^{-10} eV $ ($\epsilon_F/\hbar
\approx 150~ ms^{-1}$) and $k_{F(2)} \approx 2.4/\mu m$.
Note that the critical temperature of the BKT transition in the BEC regime of interest to
us is estimated to be $T_{BKT} \sim 0.1 \epsilon_{F(2)} \sim 500 {\rm nK}$~\cite{bab,mat} below which the system can still have
sizable condensates with ignorable thermal fluctuations~\cite{ries,mur}.
Assuming this, in the deep BEC regime (\ref{deltat2_s})  gives
\be
 \left(\overline{\Delta T_{(2)}} \right)^2 \approx 6 \, T_s^2 > T_s^2 \, , \label{deltat2_s2}
 \ee
on the margin of acceptability of (\ref{ggg}).

In terms of the
dimensionless coupling $\bar g_{(2)}$ defined in~(\ref{deltat2max}) , $^6Li$ at the density above
corresponds to ${\bar g}^2_{(2)} = 1.44/16 \pi$ for $g^2_{(2)} \approx (\epsilon_{F (2)}^2/k_{F(2)}^2)$ (the value of the coupling constant chosen to produce the 2D result of Fig.~\ref{deltaT}).
Then $g^2_{(3)}=g^2_{(2)} L_{z} \approx 10  (\epsilon^2_{F (3)}/
k_{F (3)}^3)$ with $k_{F (2)}= \sqrt{2 k_{F (3)} L_z/ 3\pi} k_{F (3)}$, giving the dimensionless coupling constant $\bar g^2_{(3)}\approx 0.8$, as
required for the narrowness  of the resonance. {Further, from (\ref{c2approx}) we have
\be
c_2\approx\frac{v_{F(2)}}{16\pi u_2^2}.
\label{c2approx2}
\ee
}

The main figure in Fig.~\ref{deltaT} shows the behaviour of $\overline{\Delta T_{(2)}}$ for
the value of $1/a_{S (2)} k_{F (2)}=0.5$ near the crossover as determined from
Eq.(\ref{deltat/t_exact}).
In the inset (bottom) to Fig.~\ref{deltaT} we plot
$\left(\epsilon_{F(2)}\overline{\Delta T_{(2)}} \right)^2$ obtained
from~(\ref{deltat/t_exact}) on varying $1/k_{F(2)} a_{S(2)}$. We see that (\ref{deltat2max}) is a good approximation across the medium to deep BEC regime and we do not have to be too careful where we are in the BEC regime to get essentially the same delay.
 The maximum travel time fluctuation occurs when $1/a_{S(2)} k_{F(2)} >
1.0$ in the BEC regime.
  If we tune the system to $1/a_{S(2)} k_{F(2)} =1.0$ the
central momentum  of the density waves is $k_0 \approx 0.02 k_{F(2)}$, determined from (\ref{c2approx2}) as
$c_2 \approx 0.02 v_{F(2)}$ at $1/a_{S(2)} k_{F(2)} =1.0$ in Fig.1.
The range of wavenumber $k$ of density perturbations is required to be $k< k_0+K$ for those modes having the same sound speed where
$K$ is found to be $K \ge k_0$, and with $k_{F(2)}
\approx 1.0 /\mu m $,  the width of the density fluctuations can be of
order 20 $\mu m$.

For
$c_2\approx 0.02 v_{F(2)} \approx 18 \, \mu m/ms$,  the travel time of the density waves over the condensate size $L_\bot  \approx 480 \mu m$ is
approximately $27 ms$~\cite{joseph}, { satisfying the first inequality of (\ref{ggg}) easily.} Additionally, the  correlation length $l=1/m c\approx 100 \mu m \gg L_z=1.4 \mu m$ so that the quasi-2D approximation can be justified.
From Fig. 1, $\overline{\Delta T_{(2)}} \approx 1.7 \epsilon_{F(2)}^{-1} \approx 0.01 ms$ at $1/a_{S(2)} k_{F(2)}=1.0$. Unfortunately, the effect is not yet testable since
experimentalists most easily measure the (saturated) fluctuations
$\Delta r = c_2 \overline{\Delta T_{(2)}}\approx 0.2 \mu m$  in the position of the
propagating wavefront. Currently, such uncertainty is well within the
noise  by between one and two orders of
magnitude \cite{joseph}.

\section{Opening the trap}

Since the saturated value $(\overline{\Delta T_{(2)}})^2$ in~(\ref{deltat2_s}) increases with decreasing fermion density $\rho_{0(2)}$, the best chance of observing fluctuations is when   the cloud of atoms undergoes free expansion as the confining {optical} potentials are switched off~\cite{meno,Lee4} spontaneously. Anticipating that the condensate expands primarily in the $z$ direction, namely $L_{\bot} \approx L_{\bot 0}$, simple dimensional analysis suggests that the dependence of $\overline{\Delta T_{(2)}}$ on the thickness $L_z$ is
\be
(\overline{\Delta T_{(2)}})^2 \propto L_z^{1/2}/(\rho_{0(3)} g_{(3)}) \propto L_z^{3/2} \,. \label{T2_Lz}
\ee
\noindent
To arrive at (\ref{T2_Lz}) we have used the results $\epsilon_F \propto \rho_{0(2)} \propto \rho_{0 (3)}^{2/3}$ and ${\bar g}^2_{2} \propto g^2_{2} /\rho_{0 (2)} \propto g^2_{3}/(L_z \rho_{0(3)}^{2/3})$,
where $\rho_{0(3)} \propto k_{F(3)}^3\propto L_z^{-1}$.
As $L_z$ increases from its initial value $L_{z0}=1.4 \mu m$ to $L_{zt}$ at time $t$ after the opening of the trap
$(\overline{\Delta T}_{(2)})_t^2 $, the moving saturated value at time $t$, increases correspondingly.

To be more quantitative we assume that the condensate behaves as a simple fluid, beginning in the BEC regime where (\ref{deltat2max}) is valid.
\begin{figure}
\centering
\includegraphics[width=0.99\columnwidth=0.8]{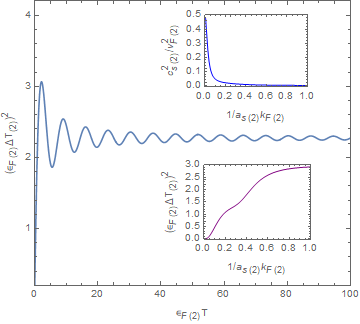}
\caption{ The Figure shows the behavior of $\left(\epsilon_{F(2)} \Delta
T \right)^2$ as a function of  $\epsilon_{F(2)} T$  at $1/a_{S(2)} k_{F(2)}=0.5$ near the crossover,
with $\bar g^2=1.44/(16\pi)$,thus $g^2 =1.44 (\epsilon_F^2/k_F^2$)%,
given by (\ref{deltat/t_exact}).  The upper inset Figure displays the result of sound speed as a function of  $1/a_{S(2)} k_{F(2)}$.
The lower
inset Figure shows the saturation value of fluctuations in
time of the flight by changing $1/a_{S(2)} k_{F(2)}$,  obtained from
(\ref{deltat2_s}).  Its maximum value occurs at $1/a_{S(2)} k_{F(2)}
> 1.0$ in the BEC regime.}
\label{deltaT}
\end{figure}
With $\delta h\propto \rho $ in the Bernoulli equation~(\ref{BLeq}), the density takes the scaling form
 \be
  \rho(x,y,z,t) = \frac{1}{\prod_j b_j}\rho_0\bigg(\frac{x}{b_x},\frac{y}{b_y},\frac{z}{b_z},t\bigg)
 \ee
\noindent
where $\rho_0$ is the initial density. The continuity and Bernoulli equations (\ref{cont}) and (\ref{BLeq}) then reduce to~\cite{meno,Lee4}
\be
{\ddot b}_i - \frac{{\omega_i^2}}{{b_i\prod_jb_j } }= 0 \,.
\ee
 For the pancake shape  harmonic trap above, with frequencies $\omega_z \gg \omega_x = \omega_y = \omega_{\bot}$, the equations for the scale parameters can be further simplified as
\be
{\ddot b}_z=  \frac{\omega_z^2}{b_z^2 } \, , \quad\quad {\ddot b}_{\bot}=  \frac{\omega_{\bot}^2}{b_z }
\label{ddotb}
\ee
where $b_x = b_y = b_{\bot}$. {With ${\ddot b}_z/{\ddot b}_{\bot} = O(10^8) $ the neglect of lateral expansion is justified.} As for $b_z(t)$, the solution for initial conditions $b_i (0)=1, \dot{b}_i (0)=0$ is
\be
b_z (t) \approx \sqrt{2} \omega_z t,
\ee
whence $b_z (T) \approx \sqrt{2} \omega_z T \approx 135$ when $T$ is a travel time about $27 ms$, appropriate to the condensate. For the parameters in~\cite{strecker} the speed of the hydrodynamic expansion $v_z={\dot b_z(t)} L_{z 0} \approx \sqrt{2} \omega_z  L_{z 0} \approx 7 \mu m/ms$ (where $L_{z 0}=1.4 \mu m$ is the initial thickness, and $\omega_z=2 \pi \times 800 Hz$) is slow in comparison to the sound speed $c_2 \approx 18 \mu m/ms$ at $u_2=1.0$. Therefore (\ref{T2_Lz}) holds true for such an adiabatic expansion. Further, since at time $t$, $L_z \propto t$, we have, on dimensional grounds
\be
(\overline{\Delta T_{(2)}})_T^2 \approx \frac{(\overline{\Delta T_{(2)}})^2}{T} \int_0^T dt~ ({\sqrt 2}\omega_z t)^{3/2}\approx \frac{3}{5}  (\overline{\Delta T})^2 \bigg(\frac{L_{zT}}{L_{z0}}\bigg)^{3/2} \,.
\label{deltat/t_approx_2d}
\ee
as long as the size of $L_{zT}$ and the correlation length $l$ are $L_{zT} \leq l$.
The static $(\overline{\Delta T_{(2)}}) \approx 0.01 ms$, obtained above for all $u_2\gtrsim 1$ in the BEC regime, and
$L_{z T}= \sqrt{2} (\omega_z T) L_{z0}$.
%\textcolor{red}{
We stress that TOF measurements are made in the non-expanding directions of the condensate. The expansion of fluctuations is due to the drop in density and not to condensate expansion {\it per se}.
%}
\\
In practice the hydrodynamic expansion has enough time to inflate the pancake-shaped condensate substantially. After travel time $ t=T \approx 27 ms$ the anisotropy ratio is about $L_{zT} /L_{\bot}  \approx (135 L_{z0})/L_{\bot 0} \approx 1/3 $ or, equivalently, since $L_{\bot 0}\approx 5 l$, $L_{zT}\approx 1.5 l$ and the system is hardly two-dimensional.
{Although we now satisfy (\ref{ggg}) unambiguously, the time $T$ for expansion is still not enough for this condensate to develop measurable fluctuations, even if we were to maintain a quasi-2D} approximation. The density reduction due to hydrodynamic expansion will only enhance the uncertainty in $\Delta r$ as estimated above to
approximately $ 30 \Delta r$ with $L_{zT}/L_{z0}=135$, say about $ 5-6 \mu m$. Although a huge improvement, this still gives a noise to signal ratio of about 2-3.

What can we say about the fluctuations in the TOF when the condensate is more 3D than we would wish? The consequence of late-time three-dimensionality on the condensate is still to increase the magnitude of the fluctuations but the effect is mixed. Increased dimensionality looks to diminish the impact of lower density since, in the BEC regime for a 3D condensate \cite{Lee3},
{
\be
{\left(\overline{\Delta T}_{(3)} \right)^2 \approx \eta^{1/2}_{(3)}/(\rho_{0 (3)})^{3/2}\propto L_{\bot}^2L_z
\label{delta/t approx5}}
\ee
}
That is, with  $L_\bot$ largely unchanged we see that, compared to $\left(\overline{\Delta T}_{(2)} \right)^2  \propto L_{z}^{3/2}$ in 2D, the 3D expansion tends to give less enhancement to  the TOF fluctuations. Further, whereas $(\overline{\Delta T}_{(2)})^2$ is largely insensitive to inverse scattering length $u_2$ within the BEC regime, $(\Delta T_{(3)})^2$ is not, vanishing in the deep BEC regime with $\eta_{(3)}^{1/2}$ \cite{Lee3}, although there is a strong peak in the early BEC regime when $u_3\sim 1$ for which (as in the D=2 case), $\epsilon_F\overline{\Delta T}_{(3)}\sim 1$. Even if we keep close to the peak the effect of the 3D regime is small. It can be shown that, for the case in hand, on assuming a rapid switch from 2D to 3D, the 3D regime only gives a few percent increase to $\overline{\Delta T_{(2)}}$ and we ignore it.

Thus, in order to make the 2D fluctuations truly visible, we need bigger condensates (to increase $T$) and/or higher frequency traps or, more simply, a larger aspect ratio of the initial pancake condensate.
There is some hope in a more recent experiment ~\cite{ries} (although not for our narrow resonance in $^6Li$), which shows how TOF fluctuation measurements could be made larger.  We take as a guide their initial trap potential frequency in the $z$-direction of order $ 8 kHz$ for a thickness of about $L_{z 0}=0.1 \mu m$  while keeping $\omega_{\bot} =2 \pi \times 0.05 Hz$ ($L_{\bot}\approx 480 \mu m$) unchanged  with the same
 2D density, $\rho_{0 (2)} \sim 10^8 (cm)^{-2}$ as above.
 The expansion speed of the cloud of  atoms after removing the trap is  $16 \mu m/ms$, about  the same order of the sound speed. Thus Eq.(\ref{T2_Lz}), strictly speaking based upon the quasi-static approximation, barely holds. Nonetheless, taken as it stands, with this initial trap potential frequency the effect of free expansion  can make the scale parameter as large as $b_z (T=27 ms) \approx 1350$, giving an anisotropy ratio that is more 2-dimensional with {$ L_{zT} \approx (1350 L_{z0}) \approx 1.35 l$ (since $l\approx 100\mu m$ is unchanged).}
 \\
On restricting ourselves to this  quasi-2D part of the expansion
the contribution to $\overline{\Delta T_{(2)}}$ is, from (\ref{deltat/t_approx_2d})
 $(\Delta T_{(2)})_{T} \simeq 140 (\overline{\Delta T_{(2)}})$ where $\overline{\Delta T_{(2)}} =0.01 ms$.
This result, that the expansion leads to two order-of-magnitude enhancement in $\Delta r = c\overline{\Delta T}$
to $140 \Delta r\approx 25 \mu m$, is potentially measurable~\cite{joseph} with a signal to noise ratio of 2.0.

\section{Conclusions}
The aim of this paper has been to see if the collapse in density of a trapped condensate of ultracold fermi atoms on releasing the trap is sufficient to give observable quantum fluctuations in phonon times-of-flight (TOF) as a result of quantum fluctuations of the sound-cone. We had shown earlier \cite{Lee3} that, for a static trapped condensate, the fluctuations were too small to be seen but, with the decrease in density that releasing the trap permits, together with reducing the dimensionality of the condensate, the fluctuations could grow to observable size.

It is clear that there are many caveats with our idealised model of a pancake $^6Li$ condensate controlled by a narrow resonance {in an optical trap}, in which we ignore edge effects and have difficulty in maintaining two-dimensionality. Also,
any finite temperature effects in the experiment will bring in thermal fluctuations that potentially will blur the signals although we hope that the rapid expansion gives mitigating cooling. 
%\textcolor{red}
{We stress again that we are not interested in fluctuations of the metric {\it per se}, but only in purely quantum fluctuations for reasons given in the introductory section.}
Accepting the model as it stands we see that, with currently accessible experimental parameters~\cite{ries}, releasing the trap can give two orders of magnitude enhancement on the hitherto unobservable static TOF fluctuations. This makes intrinsically quantum mechanical TOF fluctuations potentially measurable~\cite{joseph} with a signal to noise  ratio greater than unity  and the possibility of even more visible fluctuations with larger and thinner condensates.

{\it Acknowledgements.---}This work was supported in part by the
Ministry of Science and Technology, Taiwan.

\appendix
\section{Appendix . The gap equation and coefficients in 2D}\label{append1}

The action $S_{NL}$ is invariant under global $U(1)$ transformations of $\varphi$ but the variational equation $\delta S_{NL} = 0$ permits constant condensate solutions $\varphi (x) = \varphi_0$, which break this symmetry spontaneously, determined by the gap equation
\begin{equation}
-\frac{N_0}{k_F {a}_S} = \frac{1}{U} = \int^{\Lambda}
\frac{ d^{2} { p} }{ (2 \pi)^2 } \frac{1}{2E_{p}} \,. \label{aS}
\end{equation}
In (\ref{aS}) $U \equiv g^2/(\nu - 2\mu)$, $E_{p}=(
\varepsilon_{p}^2 + g^2|\varphi_{0}|^2 )^{1/2} $ and the 2D scattering length $a_S$ is determined
from the 3D scattering length and the spatial confinement length scale $L_z$ (see~\cite{a1,a2,a3,a4,a5} ).
The equation is
UV-singular and regularization/renormalization is
needed. Moreover, it is known that in two dimensions a
bound state of pairs of fermi atoms exists, energy $ E_B=1/ m {a}_S^2$,
 for any coupling strength~\cite{a1}, where
\be
 \frac{1}{{U}} = \int^{\Lambda} \frac{
d^{2} { p} }{ (2 \pi)^2 } \frac{1}{ |E_B|^2+{\bf p}^2/m} \,.
\label{bound_state}
\ee
A finite equation for
  the condensate
$ \varphi_{0}$ for a given scattering length
${a}_S$ (and for the chemical potential $\mu$) can be found by subtracting
equation~(\ref{bound_state}) from
equation~(\ref{aS}), giving
\be
0=\int \frac{ d^{2} { p} }{ (2 \pi)^2 } \bigg(\frac{1}{ 2E_{\bf
p}}-\frac{1}{ |E_B|^2+{\bf p}^2/m} \bigg) \, ,
\label{subtracted_gapeq}
\ee
which can be cast as
$
\sqrt{\mu^2+{g}^2 |\varphi_{0}|^2}=E_B +\mu \,.
$
The {\it constant} 2D fermion number density  $\rho_0 =k_F^2/2 \pi$ is decomposable as $\rho_0 = \rho^F_0 +\rho^B_0 $, where $\rho^F$ is the explicit fermion density and $\rho^B$
is the molecule density (two fermions per
molecule). It is sufficient to give
\begin{eqnarray}
\rho^F_0 &&=\frac{m}{ 2\pi} \big[ \mu +\sqrt{\mu^2+ g^2 \varphi_0^2} \big] \nonumber\\
&& =  \frac{m \epsilon_F}{ 2\pi}
\bigg[ 1+ \sqrt{1+ 4 \bar g^2 \big(\frac{u^2}{\bar g^2+ u^2}\big)^2/\big(u^2-\frac{\bar g^2}{ \bar g^2+u^2}\big)^2 }\bigg] \nonumber\\
\rho^B_0 &&=
  2|\varphi_0|^2 \!\!= \frac{m \epsilon_F}{\pi} \bigg(\frac{u^2}{\bar{g}^2+u^2} \bigg) .
   \label{rho0}
  \end{eqnarray}
\noindent
In (\ref{rho0}) the dimensionless coupling constant $\bar{g}$ is defined by  $g^2=( 8 \pi \epsilon_F/m) \bar{g}^2$ and  $u =1/{a_S k_F}$.
As expected, in the deep BCS regime $u \ll 1$, $ \vert \varphi_0 \vert $ is vanishingly small with  $\rho_0^F \gg \rho_0^B$,  while in the deep BEC regime $u \gg 1$  $\vert \varphi_0 \vert^2 \rightarrow \rho_0/2$
with $\rho_0^B \gg \rho_0^F$.

The coefficients of (\ref{LeffU0}) can be expressed by elementary functions as
\ba
\label{phimu}
&& \!\!\!\!\mu \!=\! \epsilon_F \bigg( \frac{ \bar{g}^2}{ \bar{g}^2 + u^2 } - u^2 \bigg), |\varphi_0|^2 \!\!= \frac{m \epsilon_F}{2\pi} \bigg(\frac{u^2}{\bar{g}^2+u^2} \bigg)
\ea
%\vskip -0.2cm
%
\begin{widetext}
\begin{eqnarray}
N_{0} &=& {g}^2 |{\varphi}_0|^2\int \frac{d^{2} {
p}}{(2\pi)^2} \frac{1}{ 2 E_{p}^3}
=\frac{m}{4\pi} \bigg[
1+\frac{\mu}{\sqrt{\mu^2+{g}^2 |{\varphi}_0|^2 }}\bigg]
=\frac{m}{2\pi} \bigg[ \frac{1}{1+u^2 + u^4/\bar{g}^2} \bigg] \, ,
\\
\alpha/\sqrt{\kappa}&=& 2|{\varphi}_{0}| \bigg[1 +
\frac{1}{2}{g}^2\int \frac{d^2 { p}}{(2\pi)^2}
\frac{\varepsilon_{p}}{ 2E_{p}^3}\bigg]
=2|{\varphi}_{0}|
 \bigg[
1+\frac{{g}^2}{8\pi}\frac{m}{\sqrt{\mu^2+{g}^2
|{\varphi}_0|^2}}\bigg]
= 2|{\varphi}_{0}|
 \bigg[ 1+\frac{\bar{g}^2 + u^2}{1+ u^2+u^4/\bar{g}^2}\bigg]\, ,
\\
{\bar M}^2/\kappa &=& 2  {g}^2 \int \frac{d^2 { p}}{
(2\pi)^2} \bigg[\frac{1}{E_{p}}-{\frac{\varepsilon_{p}^2}{ E_{p}^3}
}\bigg]
= \frac{m}{\pi} {g}^2  \bigg[
1+\frac{\mu}{\sqrt{\mu^2+{g}^2 |{\varphi}_0|^2}}\bigg]
= \frac{ 16 \, \epsilon_F \, \bar{g}^2}{ 1+u^2+u^4/\bar{g}^2} \, ,
\\
\eta/\kappa &=&  {g}^2  \int \frac{d^2 { p}}{(2\pi)^2}
\frac{\varepsilon_{p}^2}{ 2E_{p}^5}\,
 = \frac{  m}{12 \pi
{g}^2 |{\varphi}_0|^2} {g}^2  \bigg[
1+\frac{\mu^3}{\big(\sqrt{\mu^2+{g}^2
|{\varphi}_0|^2}\big)^3}\bigg]
= \frac{1}{3 \epsilon_F} \frac{(\bar{g}^2+u^2) \, ( 1+3 u^4 (1+u^2/\bar{g}^2)^2)}{u^2 \, (1+u^2(1+ u^2/\bar{g}^2))^3} \, .\label{defs}
\end{eqnarray}
The scale factor is
\be \kappa \;=\;
\frac{\rho_0}{4m {g}^2\zeta + 2} \, ,
\label{kappa}
\ee
where
\ba
\zeta &=&  \int \frac{d^2 { p}}{ (2\pi)^2}\bigg[\frac{1}{8
E_{p}^3} \bigg[ \bigg( 1-3 \frac{{g}^2
|\varphi_0|^2}{E_{p}^2}\bigg) \frac{\varepsilon_{p}}{m}  + \bigg(
5 \frac{{g}^2 |\varphi_0|^2}{E_{p}^2} \bigg(
1-\frac{{g}^2 |\varphi_0|^2}{E_{p}^2}\bigg) \bigg)
\frac{|{\bf p}|^2 \, ({\hat {\bf p}} \cdot {\hat \nabla})^2 }{m^2}
\bigg] \bigg] \,
\nonumber
\\
&=& \frac{1}{24 \pi} \frac{\mu^4+ 3 \mu^2 g^2 \vert \varphi_0\vert^2+ g^4 \vert \varphi_0\vert^4+ \mu \big(\sqrt{\mu^2+g^2 \vert\varphi_0\vert^2}\big)^3 \,}
{g^2 \vert \varphi_0\vert^2 \, \big(\sqrt{\mu^2+ g^2 \vert\varphi_0\vert^2}\big)^3} \nonumber
\\
&=&
{
\frac{1}{ 96 \pi \epsilon_F} \frac{(\bar g^2+u^2 (\bar g^2+x^2))^4 + 4 \bar g^2 u^2 (\bar g^2-u^2 (\bar g^2+u^2))^2+ (\bar g^2-u^2 (\bar g^2+u^2)) (\bar g^2+u^2 (\bar g^2+u^2))^3}{\bar g^2 u^2 (\bar g^2+u^2 (\bar g^2+u^2))^3}} \, . \nonumber
\\
\label{zeta}
\ea
In (\ref{zeta}) $ {\hat {\bf p}}$ and  ${\hat \nabla}$ are the unit
vectors along the direction ${\bf p}$ and the direction of the
spatial variation of the phase mode $\theta$ respectively.
\\

The phonon has dispersion relation
 $\omega^2 = c^2 {\bf k}^2$,
 with speed of sound
\be
c^2=\frac{\rho_0/2m}{N_0+\frac{{4\alpha^2}}{\bar{M}^2}}=\frac{v_F^2}{2} \frac{\bar{g}^2 \, (1+u^2+ u^4/\bar{g}^2)\, (u^2+\bar{g}^2)}{ \bar{g}^2 (u^2+\bar{g}^2)+u^2 (1+ \bar{g}^2+ 2 u^2+ u^4/\bar{g}^2)^2 } \,.
\label{c2}
\ee
In the deep BCS regime $u\ll 1$, $c^2 \rightarrow v_F^2/
2$, as expected, and in the deep BEC regime $u \gg 1$, $c \rightarrow 0$
\end{widetext}

\end{document}